\documentclass{appolb}
\usepackage{amssymb,amsmath}
\usepackage{epsfig}

\begin{document}

\title{Citation entropy and research impact estimation}

\author{Z.~K.~Silagadze 
\address{Budker Institute of Nuclear Physics \and  
Novosibirsk State University \\ 630 090, Novosibirsk, Russia }}

\maketitle

\begin{abstract}
A new indicator, a real valued $s$-index, is suggested to characterize a
quality and impact of the scientific research output. It is expected to be 
at least as  useful as the notorious $h$-index, at the same time avoiding 
some its obvious drawbacks. However, surprisingly, the $h$-index is found 
to be quite a good indicator for majority of real-life citation data with 
their alleged Zipfian behaviour for which these drawbacks do not show up.

The style of the paper was chosen deliberately somewhat frivolous to indicate 
that any attempt to characterize the scientific output of a researcher by 
just one number always has an element of a grotesque game in it and should 
not be taken too seriously. I hope this frivolous style will be perceived as 
a funny decoration only.
\end{abstract}

\section*{}
Sound, sound  your trumpets and beat your drums! Here it is, an impossible 
thing performed: a single integer number characterizes both productivity and 
quality of a scientific research output. Suggested by Jorge Hirsch \cite{1},
this simple and intuitively appealing $h$-index has shaken academia like
a storm, generating a huge public interest and a number of discussions and
generalizations \cite{2,3,4,5,6,7,8,9,10,11,11P}.

A Russian physicist with whom I was acquainted long ago used to say that the 
academia is not a Christian environment. It is a pagan one, with its 
hero-worship tradition. But hero-worshiping requires ranking. And a simple 
indicator, as simple as to be understandable even by dummies, is an ideal
instrument for such a ranking.

$h$-index is defined as given by the highest number of papers, $h$, which 
has received $h$ or more citations. Empirically, 
\begin{equation}
h\approx\sqrt{\frac{C_{tot}}{a}},
\label{hind}
\end{equation}
with $a$ ranging between three and five \cite{1}. Here $C_{tot}$ stands for
the total number of citations.

And now, with this simple and adorable instrument of ranking on the pedestal, 
I'm going into a risky business to suggest an alternative to it. Am I 
reckless? Not quite. I know a magic word which should impress pagans with an 
irresistible witchery. 

Claude Shannon introduced the quantity 
\begin{equation}
S=-\sum\limits_{i=1}^N p_i\ln{p_i},
\label{Sentropy}
\end{equation}
which is a measure of information uncertainty and plays a central role in 
information theory \cite{12}. On the advice of John Von Neumann, Shannon
called it entropy. According to Feynman \cite{13}, Von Neumann declared to 
Shannon that this magic word would give him "a great edge in debates because 
nobody really knows what entropy is anyway".

Armed with this magic word, entropy, we have some chance to overthrow the
present idol. So, let us try it! Citation entropy is naturally defined by 
(\ref{Sentropy}), with
$$p_i=\frac{C_i}{C_{tot}},$$
where $C_i$ is the number of citations on the $i$-th paper of the citation 
record. Now, in analogy with (\ref{hind}), we can define the citation record
strength index, or $s$-index, as follows
\begin{equation}
s=\frac{1}{4}\sqrt{C_{tot}}\;e^{S/S_0},
\label{sind}
\end{equation}
where
$$S_0=\ln{N}$$
is the maximum possible entropy for a citation record with $N$ papers in 
total, corresponding to the uniform citation record with $p_i=1/N$.

Note that (\ref{sind}) can be rewritten as follows
\begin{equation}
s=\frac{1}{4}\sqrt{C_{tot}}\;e^{(1-S_{KL}/{S_0})}\approx
\frac{2}{3}\sqrt{C_{tot}}\;e^{-S_{KL}/{S_0}},
\label{sind1}
\end{equation}
where
\begin{equation}
S_{KL}=\sum\limits_{i=1}^N p_i\ln{\frac{p_i}{q_i}},\;\;q_i=1/N,
\label{KLentropy}
\end{equation}
is the so called Kullback-Leibler relative entropy \cite{14}, widely used
concept in information theory \cite{15,16,17}. For our case, it measures the 
difference between the probability distribution $p_i$ and the uniform 
distribution $q_i=1/N$. The Kullback-Leibler relative entropy is always 
a non-negative number and vanishes only if $p_i$ and $q_i$  probability 
distributions coincide.   

That's all. Here it is, a new index $s$ afore of you. Concept is clear and 
the definition simple. But can it compete with the $h$-index which already 
has gained impetus? I do not know. In fact, it does not matter much whether 
the new index will be embraced with delight or will be coldly rejected with 
eyes wide shut. I sound my lonely trumpet in the dark trying to relax at the 
edge of precipice which once again faces me. Nevertheless, I feel $s$-index 
gives more fair  ranking than $h$-index, at least in the situations 
considered below.

Some obvious drawbacks of the $h$-index which are absent in the suggested
index are the following:
\begin{itemize}
\item $h$-index does not depend on the extra citation numbers of papers which
already have $h$ or more citations. Increasing the citation numbers of most
cited $h$ papers by an order of magnitude does not change $h$-index. Compare, 
for example, the citation records $10,10,10,10,10,10,10,$ $10,10,10$ and
$100,100,100,100,100,100,100,100,100,100$ which have $h=10,s=6.8$ and
$h=10,s=21.5$ respectively.
\item $h$-index will not change if the scientist losses impact (ceases to be
a member of highly cited collaboration).  For example, citation records
$10,10,10,10,10$ and $10,10,10,10,10,0,0,0,0,0,0,0,0,0,0,0,0,0,0,0$ both have 
$h=5$, while $s$-index drops from 4.8 to 3.0.
\item $h$-index will not change if the citation numbers of not most cited 
papers increase considerably. For example, $10,10,10,10,10,0,0,0,0,0,0,$ 
$0,0,0,0,0,0,0,0,0$ and $10,10,10,10,10,4,4,4,4,4,4,4,4,4,4,4,4,4,$\\ $4,4$ 
both have $h=5$, while $s$-index increases from 3.0 to 6.9.
\end{itemize}

Of course, $s$-index itself also has its obvious drawbacks. For example,
it is a common case that an author publishes a new article which gains at the 
beginning no citations. In this case the entropy will typically decrease 
and so will the $s$-index. I admit such a feature is somewhat 
counter-intuitive for a quantity assumed to measure  the impact of scientific 
research output. However, the effect is only sizable for very short citation
records and in this case we can say that there really exists some amount of 
objective uncertainty in the estimation of the impact.

Anyway, you can hardly expect that a simple number can substitute for 
complex judgments implied by traditional peer review. Of course, the latter
is subjective. Nothing is perfect under the Moon. It is tempting, therefore, 
to try to overcome this possible subjectivity of peer review by using ``simple 
and objective'' numerical measures. Especially because some believe academic
managers ``don't have many skills, but they can count'' \cite{18}. In 
reality, however, it is overwhelmingly evident \cite{19} that simple 
numerical indicators, like $s$-index proposed here, neither can eliminate
subjectivity in management science nor prevent a dull academic management.
But they can add some fun to the process, if carefully used: ``citation 
statistics, impact factors, the whole paraphernalia of bibliometrics may in
some circumstances be a useful servant to us in our research. But they are 
a very poor master indeed'' \cite{20}.

\begin{figure}[htbp]
\begin{center}
\mbox{\epsfig{figure=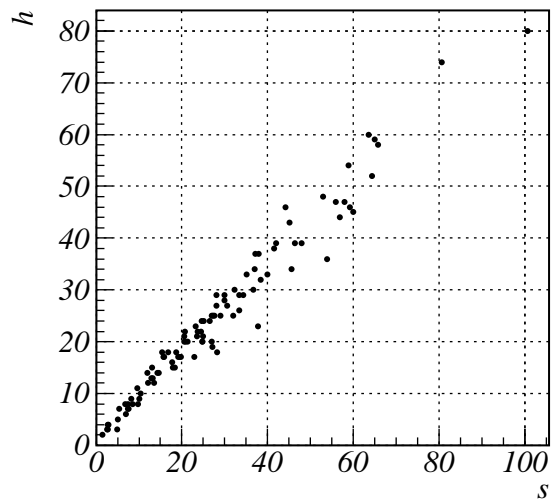}}
\end{center}
\caption {$h$-index versus $s$-index for one hundred {\em Citebase} records.}
\label{sh}
\end{figure}

It will be useful to compare two ``servants'' on some representative set of 
real-life citation records and Fig.\ref{sh} gives a possibility. One 
hundred citation records were selected more or less randomly from the 
{\em Citebase} \cite{21} citation search engine. Fig.\ref{sh} shows $h$-index 
plotted against $s$-index for these records. 

What a surprise! $h$ and $s$ indexes are strongly correlated and almost equal 
for a wide range of their values. Of course, the coefficient, $1/4$, in 
(\ref{sind}) was chosen to make these two indexes relatively equal for 
{\it some} citation records, but I have not expected them to remain close for
{\it all} citation records.

There is some mystery here. Let us try to dig up what it is. Recent studies 
indicate that from the one hand the normalized entropy does not change much
from record to record with $S/S_0\approx 0.8$ \cite{22}, and on the other 
hand, the scaling law (\ref{hind}) for the $h$-index is well satisfied with 
$a=4$ \cite{23}. These two facts and a simple calculation imply that $s$ and 
$h$ indexes are expected to be approximately equal. Therefore, the real 
question is why these regularities are observed in the citation records?

Common sense and some experience on the citation habits tell us that 
these habits are the subject of preferential attachment -- the papers that 
already are popular tend to attract more new citations than less popular
papers. It is well known that the preferential attachment can lead to power
laws \cite{24}. Namely, with regard to citations, if the citations are ranked 
in the decreasing order $C_1\ge C_2\ge \ldots C_N$ then the Zipf's law 
\cite{25} says that
\begin{equation}
C_i=\frac{C}{i^\alpha}.
\label{Zipf}
\end{equation}

Empirical studies reveal \cite{26,27} that the Zipf's law is a ubiquitous
and embarrassingly general phenomenon and in many cases $\alpha\approx 1$. 
The citation statistics also reveals it \cite{28P,28,29}. Therefore let us 
assume the simplest case, distribution (\ref{Zipf}) with $\alpha=1$. Then 
$C_h=h$ condition determines the Hirsch index
\begin{equation}
h=\sqrt{C}.
\label{hindc}
\end{equation}
And we see that, if the individual citation records really follow the Zipf
distribution $C_i=C/i$, the Hirsch index is a really good indicator as it 
determines the only relevant parameter $C$ of the distribution. In fact, the 
number of papers are finite and we have a second parameter, the total number 
of papers $N$. For sufficiently large $N$,
$$C_{tot}=\sum\limits_{i=1}^N \frac{C}{i}\approx C\int\limits_1^N \frac{dx}{x}
=C\ln{N}.$$
Therefore, from (\ref{hindc}) we get the following scaling law
\begin{equation}
h\approx h_N=\sqrt{\frac{C_{tot}}{\ln{N}}}
\label{hindN}
\end{equation}
instead of (\ref{hind}). However, $\sqrt{\ln{N}}$ varies from $1.84$ to
$2.21$ then $N$ varies from $30$ to $130$ and this explains the observations 
of \cite{23}.

Note that the relation (\ref{hindc}) was already suggested by Egghe and 
Rousseau, on the basis of Zipf's law, in \cite{30}. For other references
where the connections between Zipf's law and $h$-index are discussed see, 
for example, \cite{5,31,32}.
 
As for $s$-index, Zipfian distribution $C_i=C/i$ implies the  
probabilities (assuming $N$ is large)
$$p_i=\frac{1}{i\ln{N}}$$
and, hence, the following entropy
$$S=\frac{1}{\ln{N}}\sum\limits_{i=1}^N \frac{1}{i}\ln{(i\ln{N})}\approx
\ln{(\ln{N})}+\frac{1}{\ln{N}}\sum\limits_{i=1}^N \frac{1}{i}\ln{i}.$$
But
$$\sum\limits_{i=1}^N \frac{1}{i}\ln{i}\approx \int\limits_1^N\frac{\ln{x}}
{x}\,dx=\frac{1}{2}\ln^2{N}.$$
Therefore,
\begin{equation}
\frac{S}{S_0}\approx \frac{1}{2}+\frac{\ln{(\ln{N})}}{\ln{N}}=
\frac{\ln{(\sqrt{N}\ln{N})}}{\ln{N}}.
\label{SS0}
\end{equation}
This expression gives $0.86$ for $N=30$ and $0.82$ for $N=130$ which are 
quite close to what was found in \cite{22} (although for a small data set).

Because
$$\frac{S_{KL}}{S_0}= 1-\frac{S}{S_0}\approx\frac{1}{2}-
\frac{\ln{(\ln{N})}}{\ln{N}}$$
is small, we have the following scaling behaviour for the $s$-index from
(\ref{sind1}):
\begin{equation}
s\approx s_N= \frac{2}{3}\,\sqrt{C_{tot}}\;\frac{\ln{(\sqrt{N}\ln{N})}}
{\ln{N}}.
\label{sindN}
\end{equation}

\begin{figure}[hp]
\begin{center}
\epsfig{figure=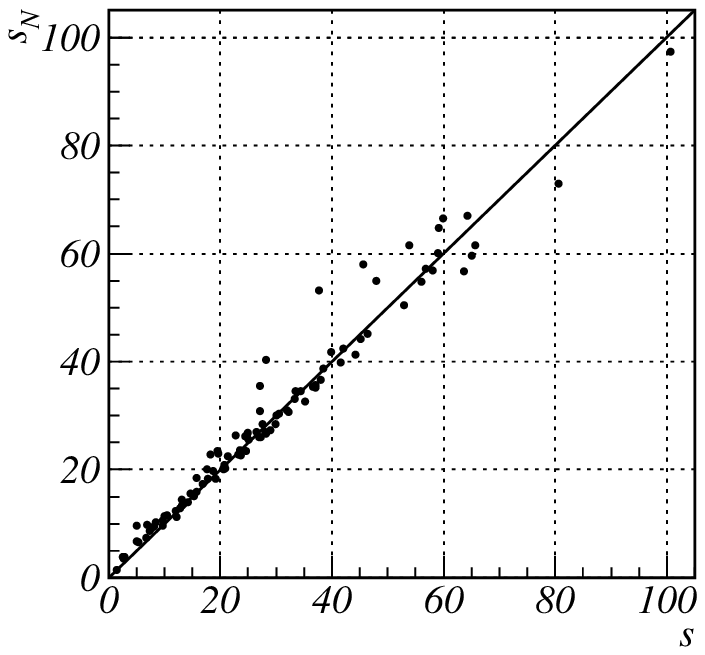}
\end{center}
\caption {$s$-index scaling: $s_N$ versus $s$-index. The solid line 
corresponds to the ideal scaling.}
\label{ssn}
\end{figure}

\begin{figure}[hp]
\begin{center}
\epsfig{figure=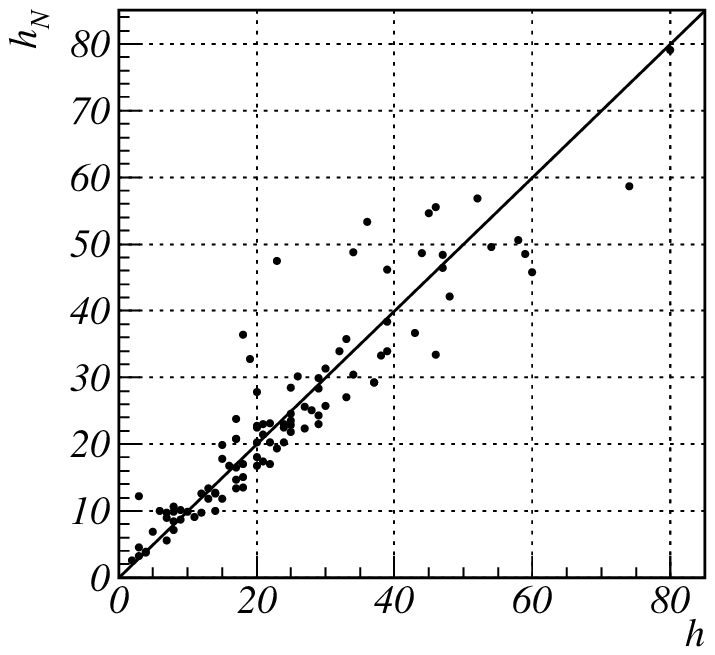}
\end{center}
\caption {$h$-index scaling: $h_N$ versus $h$-index.  The solid line 
corresponds to the ideal scaling.}
\label{hhn}
\end{figure}

Fig.\ref{ssn} and Fig.\ref{hhn} demonstrate an empirical evidence for these
scaling rules from the {\em Citebase} citation records mentioned above.
  
As we see, the scalings (\ref{sindN}) and (\ref{hindN}) are quite pronounced
in the data. However, there are small number of exceptions. Therefore, 
citation patterns are not always Zipfian. Inspection shows that in such cases
the Zipfian behavoir is spoiled by the presence of several renowned papers
with very high number of citations. If they are removed from the citation
records, the Zipfian scalings for $s$ and $h$ indexes are restored for these 
records too.
 
To conclude, we wanted to overthrow the King but the King turned out to be 
quite healthy. The secret magic that makes $h$-index healthy is Zipfian 
behaviour of citation records. Under such behaviour, a citation record really
has only one relevant parameter and the Hirsch index just gives it. However,
not all citation records exhibit Zipfian behaviour and for such exceptions 
the new index related to the entropy of the citation record probably makes 
better justice. But, I'm afraid, this is not sufficient to sound our trumpets 
and cry  ``Le Roi est mort. Vive le Roi!''

The magic has an another side however. The Zipfian character of citation
records probably indicate the prevalence of preferential attachment and
some randomness in the citation process. We can use a propagation of misprints
in scientific citations to estimate how many citers really read the original
papers and the striking answer is that probably only 20\% do, the others 
simply copying the citations \cite{33}.
 
Of course, it will go too far to assume that ``copied citations create 
renowned papers'' \cite{34}, but these observations clearly give a caveat 
against taking all the indexes based on the number of citations too 
seriously. ``Not everything that can be counted counts, and not everything 
that counts can be counted'' \cite{35}. Such local measures of the research 
impact estimation should be taken with a grain of salt and at least 
supplemented by different instruments for analyzing the whole citation 
network, like the one given in \cite{36}.
 
\section*{Acknowledgments}
The author thanks to anonymous referee for helpful and constructive comments.

\end{document}